\documentclass[aps,prl,reprint,superscriptaddress, showpacs]{revtex4-1}

\usepackage{amssymb}
\usepackage{amsmath}
\usepackage{graphicx}
\usepackage{natbib}

\begin{document}

\title{A Semiconductor Nanowire-Based Superconducting Qubit}
\author{T. W. Larsen}
\altaffiliation{These authors contributed equally to this work.}
\affiliation{Center for Quantum Devices, Niels Bohr Institute, University of Copenhagen, Copenhagen, Denmark}
\author{K. D. Petersson}
\altaffiliation{These authors contributed equally to this work.}
\affiliation{Center for Quantum Devices, Niels Bohr Institute, University of Copenhagen, Copenhagen, Denmark}
\author{F. Kuemmeth}
\affiliation{Center for Quantum Devices, Niels Bohr Institute, University of Copenhagen, Copenhagen, Denmark}
\author{T. S. Jespersen}
\affiliation{Center for Quantum Devices, Niels Bohr Institute, University of Copenhagen, Copenhagen, Denmark}
\author{P. Krogstrup}
\affiliation{Center for Quantum Devices, Niels Bohr Institute, University of Copenhagen, Copenhagen, Denmark}
\author{J. Nyg\r{a}rd}
\affiliation{Center for Quantum Devices, Niels Bohr Institute, University of Copenhagen, Copenhagen, Denmark}
\affiliation{Nano-Science Center, Niels Bohr Institute, University of Copenhagen, Copenhagen, Denmark}
\author{C. M. Marcus}
\affiliation{Center for Quantum Devices, Niels Bohr Institute, University of Copenhagen, Copenhagen, Denmark}

\begin{abstract}
We introduce a hybrid qubit based on a semiconductor nanowire with an epitaxially grown superconductor layer. Josephson energy of the transmon-like device (`gatemon') is controlled by an electrostatic gate that depletes carriers in a semiconducting weak link region. Strong coupling to an on-chip microwave cavity and coherent qubit control via gate voltage pulses is demonstrated, yielding reasonably long relaxation times ($\sim$0.8 $\mu$s) and dephasing times ($\sim$1 $\mu$s), exceeding gate operation times by two orders of magnitude, in these first-generation devices. Because qubit control relies on voltages rather than fluxes, dissipation in resistive control lines is reduced, screening reduces crosstalk, and the absence of flux control allows operation in a magnetic field, relevant for topological quantum information.
\end{abstract}


\maketitle

Superconducting qubits present a scalable solid state approach to building a quantum information processor \cite{Devoret:2013jz}. Recent superconducting qubit experiments have demonstrated single and two-qubit gate operations with fidelities exceeding 99\%, placing fault tolerant quantum computation schemes within reach \cite{Barends:2014fu}. While there are many different implementations of superconducting qubits \cite{Bylander:2011wc, Kim:2011uj, Manucharyan:2012wm}, the key element is the Josephson junction (JJ), a weak link between superconducting electrodes. The JJ provides the necessary nonlinearity for non-degenerate energy level spacings, allowing the lowest two levels to define the qubit $|0\rangle$ and $|1\rangle$ states. High quality JJs for superconducting qubits are commonly fabricated using an insulating Al$_2$O$_3$ tunnel barrier between superconducting electrodes \cite{Paik:2011hd}. For such superconductor-insulator-superconductor (SIS) JJs, the maximum allowed supercurrent, the critical current, $I_\mathrm{c}$, and the Josephson coupling energy, $E_\mathrm{J} = \hbar I_\mathrm{c}/2 e$, where $e$ is the electron charge, are fixed and determined through fabrication. 

\begin{figure}
\begin{center}
		\includegraphics[width=1\columnwidth]{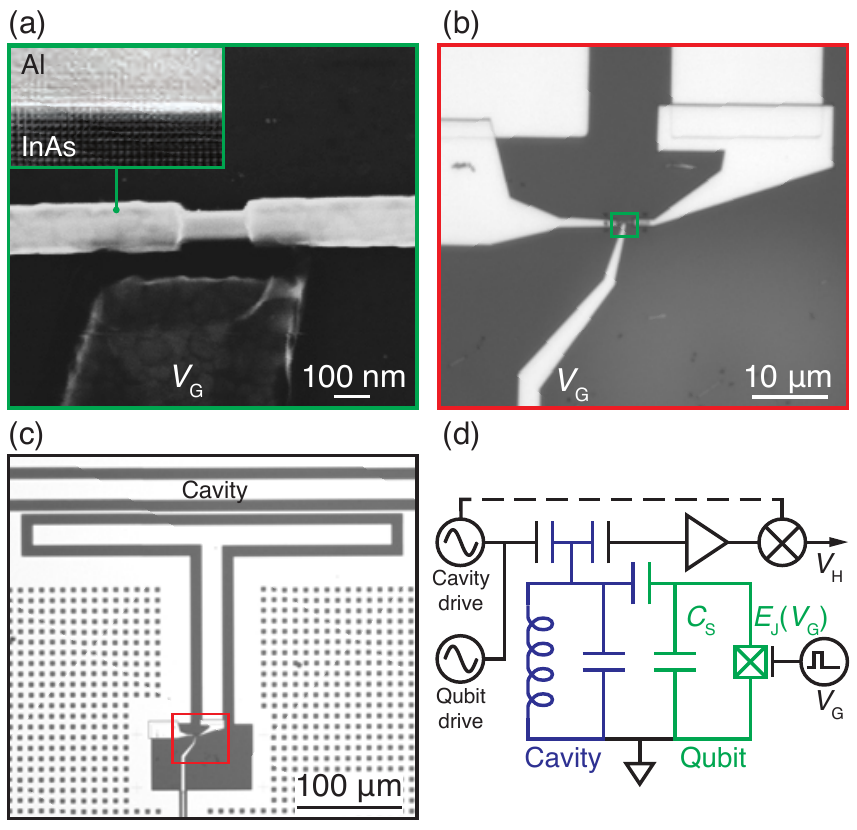}
\caption{\label{fig1} InAs nanowire-based superconducting transmon qubit. (a) Scanning electron micrograph of the InAs-Al JJ. A segment of the epitaxial Al shell is etched to create a semiconducting weak link. Inset shows a transmission electron micrograph of the epitaxial InAs/Al interface. (b)-(c) Optical micrographs of the completed gatemon device. The nanowire JJ is shunted by the capacitance of the T-shaped island to the surrounding ground plane. The center pin of the coupled transmission line cavity is indicated in (c). (d) Schematic of the readout and control circuit.}
\end{center}	
\vspace{-0.5cm}
\end{figure}

Previous work has demonstrated superconductor-normal-superconductor (SNS) JJs where the normal element is a semiconductor \cite{Doh:878391, Abay:2014ed}. Introducing a semiconductor allows $E_\mathrm{J}$ to be tuned by an electric field which controls the carrier density of the normal region and thus the coupling of the superconductors. InAs nanowires allow for high quality field effect JJs due to the highly transparent Schottky barrier-free SN interface. The recent development of InAs nanowires with epitaxially-grown Al contacts yields an atomically precise SN interface and extends the paradigm of nanoscale bottom-up technology for superconducting JJ-based devices \cite{Krogstrup:2014va, Chang:2014vf, Shim:2014}. Here we present a superconducting transmon qubit on the epitaxial InAs-Al nanowire JJ \cite{Koch:2007gz, Houck:2007dy}. This gate tunable transmon - or `gatemon' - is simply controlled using an electrostatic gate and, for this first generation of devices, shows coherence times of order 1 $\mu$s. Our results highlight the potential of using bottom-up fabrication techniques to form high quality JJ-based qubits that offer new means of electrical control. Independent research paralleling our own reports spectroscopic measurements
on hybrid qubits using NbTiN-contacted InAs nanowires \cite{deLange:2014}. 

We have fabricated and measured two gatemon devices, which show similar performance. Except where noted, data is from the first device. The qubit features a single InAs SNS JJ shunted by a capacitance, $C_\mathrm{S}$ \cite{Koch:2007gz, Houck:2007dy, Barends:2013kz}. The JJ is formed from a molecular beam epitaxy-grown InAs nanowire, $\sim$75 nm in diameter, with an \textit{in situ} grown $\sim$30 nm thick Al shell. The Al shell forms an atomically precise SN interface leading to a proximity induced gap in the InAs core with a low density of states below the superconducting gap (hard gap) \cite{Krogstrup:2014va, Chang:2014vf}. By wet-etching away a $\sim$180 nm segment of the Al shell [Fig.\ 1(a)] a weak link in the superconducting shell is formed, creating the JJ. A supercurrent leaking through the semiconductor core links the unetched regions and determines the Josephson coupling energy, $E_\mathrm{J} (V_\mathrm{G})$, which can be tuned by changing the electron density in the semiconductor core with a nearby side gate voltage, $V_\mathrm{G}$. 

As with conventional transmons, the gatemon operates as an anharmonic LC oscillator with a nonlinear inductance provided by the JJ. The total capacitance of the gatemon qubit, $C_{\Sigma}$, is determined by the capacitance of the T-shaped Al island to the surrounding Al ground plane, as shown in Fig.\ 1(c). The gatemon operates with $E_\mathrm{J} \gg E_\mathrm{C}$, where the charging energy, $E_\mathrm{C} = e^2/2C_{\Sigma}$. In this regime, decoherence due to either low frequency charge noise on the island or quasiparticle  tunneling across the JJ is strongly suppressed. For many conducting channels in the wire, the qubit transition frequency is given by $f_\mathrm{Q} = E_{01}/h \approx \sqrt{8 E_\mathrm{C} E_\mathrm{J} (V_\mathrm{G})}/h$. The difference between $E_{01}$ and the next successive levels, $E_{12}$, is the anharmonicity, $\alpha = E_{12} - E_{01} \approx -E_\mathrm{C}$. From microwave spectroscopy of our gatemon we estimate $\alpha/h \approx -100$ MHz. 

The gatemon is coupled to a $\lambda/2$ superconducting transmission line cavity with a bare resonance frequency $f_\mathrm{C} \approx 5.96$ GHz and quality factor, $Q \sim 1500 $. The cavity is used for dispersive readout of the qubit with homodyne detection [Fig.\ 1(d)] \cite{Homodyne}. Both the cavity and qubit leads are patterned by wet etching an Al film on an oxidised high resistivity Si substrate. The nanowire contacts and gate are also patterned from Al using a lift-off process with an ion mill step to remove the native Al$_2$O$_3$ prior to deposition.

\begin{figure}
\begin{center}
		\includegraphics[width=1\columnwidth]{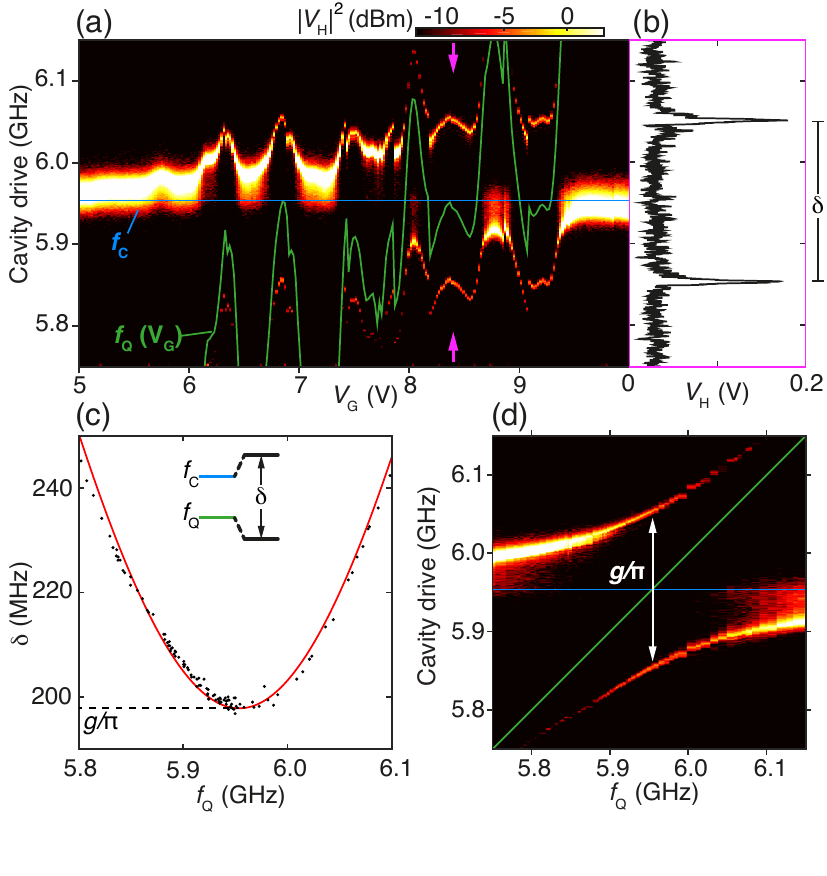}
\caption{\label{fig2} Strong coupling of the gatemon to the microwave cavity. (a) Cavity transmission as a function of the cavity drive frequency and $V_\mathrm{G}$. The solid blue line shows the bare cavity resonance frequency, $f_\mathrm{C}$, while the solid green line indicates the gate-voltage dependent qubit frequency, $f_\mathrm{Q} (V_\mathrm{G})$, extracted from the data. (b) Cavity transmission as a function of the cavity drive at the position indicated by the purple arrows in (a). (c) Frequency splitting between the hybridized qubit-cavity states, $\delta$, as a function of $f_\mathrm{Q}$, as extracted from (a). From fitting the solid theory curve we extract the qubit-cavity coupling strength, $g/2\pi = 99$ MHz. (d) Parametric plot of the data from (a) as a function of the cavity drive and qubit frequency, $f_\mathrm{Q}$.}
\end{center}	
\vspace{-0.5cm}
\end{figure}

Gatemon-cavity coupling was investigated by measuring cavity transmission at low drive power as a function of the cavity drive frequency and gate voltage, $V_\mathrm{G}$, with $f_\mathrm{Q} \sim f_\mathrm{C}$ [Fig.\ 2(a)]. Aperiodic fluctuations in the resonance as a function of $V_\mathrm{G}$, with regions of widely split transmission peaks, were observed [Fig.\ 2(b)]. These gate-dependent, repeatable fluctuations in the cavity resonance are associated with mesoscopic fluctuations in the nanowire transmission---appearing also as fluctuations of normal-state conductance, $G_\mathrm{N}(V_\mathrm{G})$ \cite{Doh:878391}---which causes fluctuations in gatemon frequency, $f_\mathrm{Q} \propto \sqrt{I_\mathrm{c} (V_\mathrm{G})}$. The changing qubit frequency, in turn, pulls on the cavity resonance, resulting in the observed response. The split cavity peaks indicate hybridized qubit and cavity states in the strong coupling regime. The coupling strength, $g$ is found to exceed the qubit and cavity decoherence rates, allowing the vacuum Rabi splitting to be resolved \cite{Wallraff:2004vt}. Writing the hybridized qubit-cavity state frequencies as $\lambda_{\pm} = \left(f_\mathrm{Q} + f_\mathrm{C} \pm \sqrt{(f_\mathrm{Q} - f_\mathrm{C})^2 + 4(g/2\pi)^2}\right)/2$, Fig.\ 2(c) shows the splitting $\delta = \lambda_{+} - \lambda_{-}$ as a function of the qubit frequency $f_\mathrm{Q}$. From the fit to the data we extract $g/2\pi = 99$ MHz. A parametric plot [Fig.\ 2(d)] of the data in Fig.\ 2(a), as a function of the extracted $f_\mathrm{Q}$, reveals the avoided crossing for the hybridized qubit-cavity states  \cite{Wallraff:2004vt}.

\begin{figure*}
\begin{center}
	\includegraphics[width=2\columnwidth]{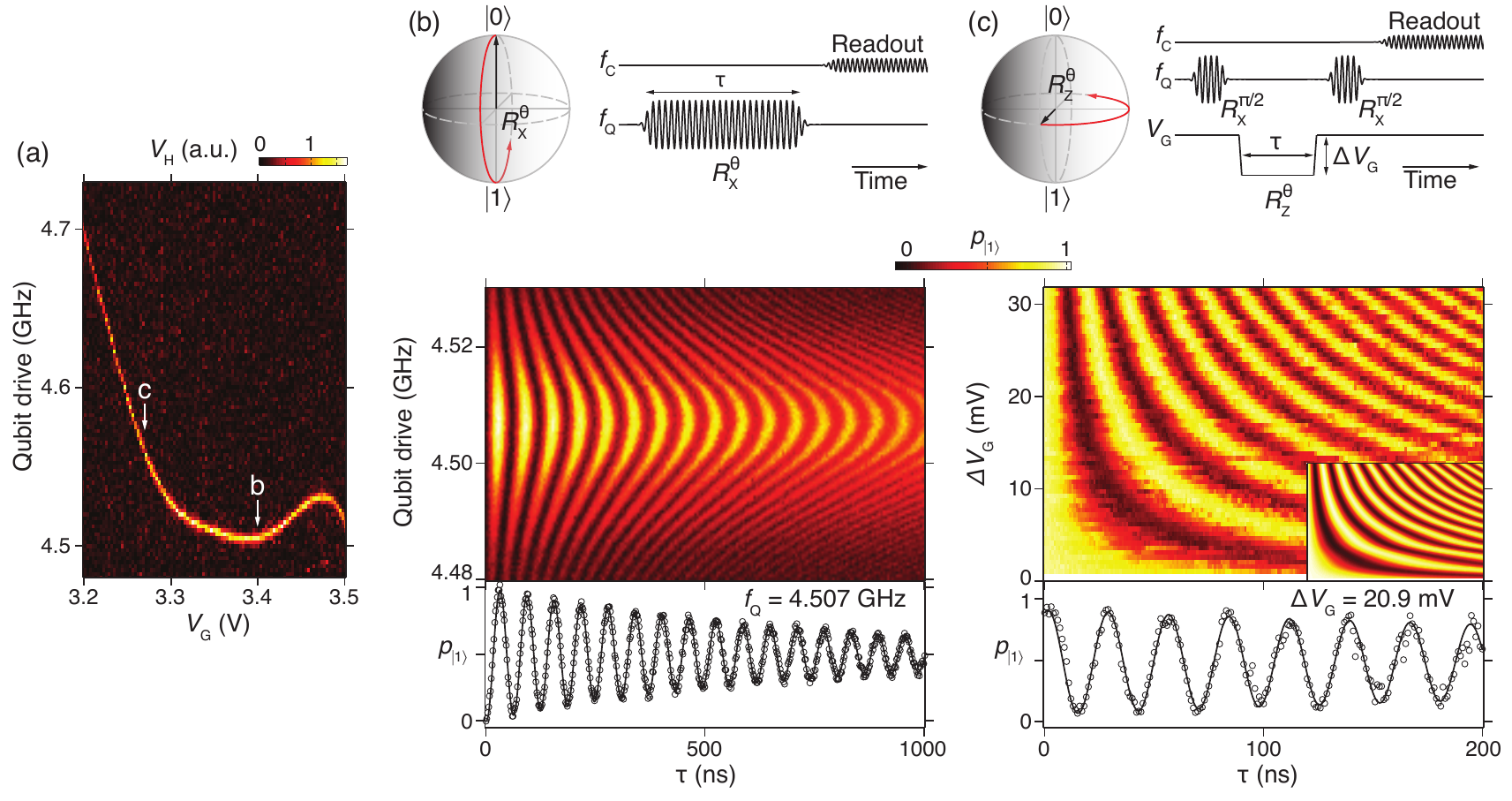}
\caption{\label{fig3} Gatemon spectroscopy and coherent control. (a) The qubit resonance frequency as a function of gate voltage, $V_\mathrm{G}$, is observed as a distinct feature. (b) Coherent Rabi oscillations are performed at point b in (a) ($V_\mathrm{G} = 3.4$ V) by applying microwave pulse for time, $\tau$, to drive the qubit followed by a readout microwave pulse to probe the cavity response. The main panel shows coherent qubit oscillations as a function of driving frequency and $\tau$. The lower panel shows coherent oscillations at the qubit resonant frequency, corresponding to rotations about the $X$-axis of the Bloch sphere. (c) Coherent oscillations about the $Z$-axis of the Bloch sphere are performed at point c in (a) ($V_\mathrm{G} = 3.27$ V) by applying a gate voltage pulse, $\Delta V_\mathrm{G}$, to detune the qubit resonance frequency for time, $\tau$. A 15 ns $R_{\mathrm{X}}^{\pi/2}$ microwave pulse is first applied to rotate the qubit into the $X$-$Y$ plane of the Bloch sphere and, following the gate pulse, a second $R_{\mathrm{X}}^{\pi/2}$ microwave pulse is used to rotate the qubit out of the $X$-$Y$ plane for readout. The main panel shows coherent $Z$ rotations as a function of $\Delta V_\mathrm{G}$ and $\tau$. The main panel inset shows the simulated qubit evolution based on $\Delta f_\mathrm{Q}(V_\mathrm{G})$ extracted from (a). The lower panel shows coherent $Z$ oscillations as a function of $\tau$ for $\Delta V_\mathrm{G} = 20.9$ mV. In both (b) and (c) the demodulated cavity response, $V_\mathrm{H}$, is converted to a normalised qubit state probability, $p_{|1\rangle}$, by fitting $X$ rotations to a damped sinusoid of the form $V_{\mathrm{H}}^0 + \Delta V_{\mathrm{H}} \mathrm{exp}(-\tau/T_{\mathrm{Rabi}})\mathrm{sin}(\omega\tau + \theta)$ to give $p_{|1\rangle} = (V_\mathrm{H} - V_{\mathrm{H}}^0)/2\Delta V_{\mathrm{H}} + 1/2$. The solid curves in the lower panels of (b) and (c) are also fits to exponentially damped sine functions.}
\end{center}	
\vspace{-0.5cm}
\end{figure*}

Demonstrations of qubit control were performed in the dispersive regime, $|f_\mathrm{Q} - f_\mathrm{C}| \gg g/2\pi$. Figure\ 3(a) shows $f_\mathrm{Q}$ as a function of gate voltage, $V_\mathrm{G}$, obtained by measuring the qubit-state--dependent cavity response following a second 2 $\mu$s microwave tone. When the qubit drive was on resonance with $f_\mathrm{Q}$, a peak in the cavity response was observed, yielding a reproducible gate voltage dependence. At a fixed gate voltage [point b in Fig.\ 3(a)] we measure in Fig.\ 3(b) the cavity response while varying the qubit drive frequency and the length of the qubit microwave pulse to observe coherent Rabi oscillations. Data in the main panel of Fig.\ 3(b) were acquired over several hours, highlighting the stability of the device.

While pulsed microwaves allow rotations about axes in the $X$-$Y$ plane of the Bloch sphere, rotations about the $Z$-axis may be performed by adiabatically pulsing $V_\mathrm{G}$ to detune the qubit resonance frequency. Such dynamic control of the qubit frequency is important for fast two qubit gate operations where the resonant frequencies of two coupled qubits are brought close to each other \cite{DiCarlo:2009ja, Barends:2014fu}. Figure\ 3(c) shows $Z$ rotations performed by first applying an $R_{\mathrm{X}}^{\pi/2}$ pulse to rotate into the $X$-$Y$ plane of the Bloch sphere followed by a negative voltage pulse, $\Delta V_\mathrm{G}$, which causes the qubit state to precess about the $Z$-axis at the difference frequency, $\Delta f_\mathrm{Q} = f_\mathrm{Q} (V_\mathrm{G} - \Delta V_\mathrm{G}) - f_\mathrm{Q}(V_\mathrm{G})$. Finally, a second $R_{\mathrm{X}}^{\pi/2}$ pulse was applied to rotate the qubit out of the $X$-$Y$ plane and measure the resulting qubit state. The observed precession frequency is consistent with the $\Delta f_\mathrm{Q}$ predicted from the spectroscopy data in Fig.\ 3(a) [Fig.\ 3(c) main panel inset].  

\begin{figure}
\begin{center}
		\includegraphics[width=1\columnwidth]{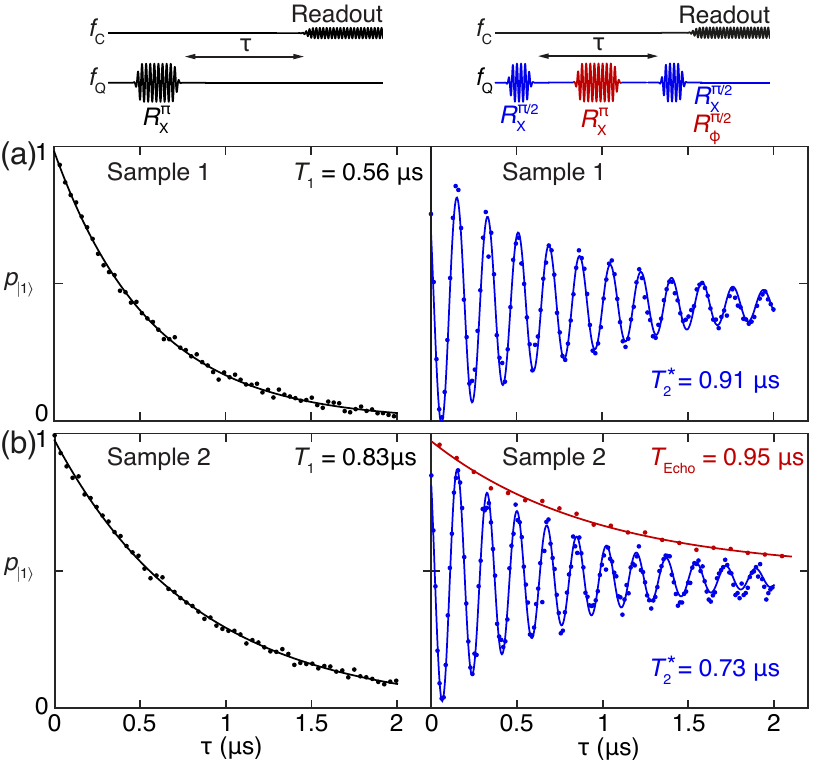}
\caption{\label{fig4} Gatemon quantum coherence. (a) Left panel shows a lifetime measurement for Sample 1 at point b in Fig.\ 3(a) ($V_\mathrm{G} = 3.4$ V). A 30 ns $R_{\mathrm{X}}^{\pi}$ pulse excites the qubit to the $|1\rangle$ state and we vary the wait time, $\tau$, before readout. The solid line is a fit to an exponential curve. The right panel shows a Ramsey experiment used to determine $T_2^*$ for Sample 1 with the wait time, $\tau$, between two slightly detuned 15 ns $R_{\mathrm{X}}^{\pi/2}$ pulses varied before readout. The solid curve is a fit to an exponentially damped sinusoid. (b) We repeat the lifetime and Ramsey experiments as in (a) for Sample 2 with $f_{\mathrm{Q}} = 4.426$ GHz ($V_\mathrm{G} = -11.3$ V). In red, we perform a Hahn echo experiment by inserting an $R_{\mathrm{X}}^{\pi}$ pulse between two $R_{\mathrm{X}}^{\pi/2}$ pulses. The decay envelope is measured by varying the phase, $\phi$, of the second $\pi/2$ microwave pulse and extracting the amplitude of the oscillations. The solid red line is a fit to an exponential curve.}
\end{center}	
\vspace{-0.6cm}
\end{figure}

Gatemon coherence times were measured quantitatively in both devices [Fig.\ 4]. The relaxation time, $T_{1}$, was measured by initializing the qubit to $|1\rangle$ and varying the waiting time, $\tau$, before readout, giving $T_1 = 0.56$ $\mu$s for the first device, measured at operating point b in Fig.\ 3(a). The decay envelope of a Ramsey measurement [Fig.\ 4(a), right panel] gives a dephasing time, $T_2^{*}  = 0.91$ $\mu$s at the same operating point. Noting that $T_2^{*} \approx 2T_1$, we conclude that at this operating point, coherence was limited by energy relaxation. Figure\ 4(b) shows coherence times for the second sample, showing a slightly longer relaxation time, $T_1 = 0.83$ $\mu$s [Fig.\ 4(b), left panel]. In this device, inhomogeneous dephasing time was shorter, $T_2^{*}  = 0.73$ $\mu$s. In Fig.\ 4(b) right panel (in red) we show that applying a Hahn echo pulse sequence, which effectively cancels low frequency noise in $f_\mathrm{Q}$, increases the dephasing time to $T_\mathrm{Echo} = 0.95$ $\mu$s. This indicates a greater degree of low frequency noise in $E_\mathrm{J}(V_\mathrm{G})$ in the second device. The observation that $T_\mathrm{Echo}$ does not reach $2 T_1$ indicates that higher frequency noise fluctuations faster than $\tau$ also contributes to dephasing. 

Coherence times for these first-generation gatemon devices are comparable to SIS transmons reported a few years ago, where typically $T_2^{*} \sim T_1 \sim 2$ $\mu$s \cite{Houck:2008je}. Longer coherence times, $T_2^{*} \sim T_1 \sim 60$ $\mu$s, have been reported more recently for planar transmon devices \cite{Chang:2013dw}. We anticipate that relaxation times can be substantially improved by removing the SiO$_2$ dielectric layer \cite{OConnell:2008jt} and more careful sample processing to reduce interface losses in the capacitor \cite{Quintana:2014jp}, and increased magnetic and infrared radiation shielding \cite{Corcoles:2011is, Barends:2011eh}. This should in turn extend dephasing times and allow for the low frequency noise spectrum to be characterized using dynamical decoupling \cite{Bylander:2011wc}. Electrical noise coupling to $E_\mathrm{J} (V_\mathrm{G})$ due to charge traps at the nanowire surface could potentially be reduced through surface passivation \cite{Ford:2012cx}. 

Conventional SIS transmons typically enable tunable frequency control by using two JJs in a SQUID geometry to create an effective flux tunable Josephson coupling energy. The qubit frequency is then controlled using superconducting current loops. The large (mA scale) currents used to control conventional flux-tuned transmons makes scaling to many qubits difficult using control electronics that pass into the cryogenic environment through normal coax lines, filters, and attenuators. On-chip voltage pulses are relatively easily screened, compared to flux pulses, which will reduce cross-talk between qubit control lines. Gatemons, with voltage tunable $f_\mathrm{Q}$, also offer new possibilities for large scale superconducting architectures. For instance, FET-based cryogenic multiplexers \cite{Ward:2013hq, AlTaie:2013ei} have recently been developed for millikelvin temperatures and would be well suited to gate control of large multi-gatemon circuits. 

Finally, we note that the epitaxial InAs-Al nanowires are expected to support Majorana bound states \cite{Mourik:2012je, Das:2012hi} due to the strong spin-orbit coupling and large $g$ factor ($\sim$10) of InAs. Recent theoretical work has proposed using transmons to manipulate and probe topologically-protected qubits built from Majorana bound states \cite{Hassler:2011gj, Ginossar:1jd}. InAs nanowire-based gatemons could therefore be readily coupled to topological qubits made using the same material technology.

%

\begin{acknowledgments}
We acknowledge financial support from Microsoft Project Q, Lundbeck Foundation, and the Danish National Research Foundation. K.D.P. was supported by a Marie Curie Fellowship.

\end{acknowledgments}


\begin{thebibliography}{31}%
\makeatletter
\providecommand \@ifxundefined [1]{%
 \@ifx{#1\undefined}
}%
\providecommand \@ifnum [1]{%
 \ifnum #1\expandafter \@firstoftwo
 \else \expandafter \@secondoftwo
 \fi
}%
\providecommand \@ifx [1]{%
 \ifx #1\expandafter \@firstoftwo
 \else \expandafter \@secondoftwo
 \fi
}%
\providecommand \natexlab [1]{#1}%
\providecommand \enquote  [1]{``#1''}%
\providecommand \bibnamefont  [1]{#1}%
\providecommand \bibfnamefont [1]{#1}%
\providecommand \citenamefont [1]{#1}%
\providecommand \href@noop [0]{\@secondoftwo}%
\providecommand \href [0]{\begingroup \@sanitize@url \@href}%
\providecommand \@href[1]{\@@startlink{#1}\@@href}%
\providecommand \@@href[1]{\endgroup#1\@@endlink}%
\providecommand \@sanitize@url [0]{\catcode `\\12\catcode `\$12\catcode
  `\&12\catcode `\#12\catcode `\^12\catcode `\_12\catcode `\%12\relax}%
\providecommand \@@startlink[1]{}%
\providecommand \@@endlink[0]{}%
\providecommand \url  [0]{\begingroup\@sanitize@url \@url }%
\providecommand \@url [1]{\endgroup\@href {#1}{\urlprefix }}%
\providecommand \urlprefix  [0]{URL }%
\providecommand \Eprint [0]{\href }%
\providecommand \doibase [0]{http://dx.doi.org/}%
\providecommand \selectlanguage [0]{\@gobble}%
\providecommand \bibinfo  [0]{\@secondoftwo}%
\providecommand \bibfield  [0]{\@secondoftwo}%
\providecommand \translation [1]{[#1]}%
\providecommand \BibitemOpen [0]{}%
\providecommand \bibitemStop [0]{}%
\providecommand \bibitemNoStop [0]{.\EOS\space}%
\providecommand \EOS [0]{\spacefactor3000\relax}%
\providecommand \BibitemShut  [1]{\csname bibitem#1\endcsname}%
\let\auto@bib@innerbib\@empty
\bibitem [{\citenamefont {Devoret}\ and\ \citenamefont
  {Schoelkopf}(2013)}]{Devoret:2013jz}%
  \BibitemOpen
  \bibfield  {author} {\bibinfo {author} {\bibfnamefont {M.~H.}\ \bibnamefont
  {Devoret}}\ and\ \bibinfo {author} {\bibfnamefont {R.~J.}\ \bibnamefont
  {Schoelkopf}},\ }\href@noop {} {\bibfield  {journal} {\bibinfo  {journal}
  {Science}\ }\textbf {\bibinfo {volume} {339}},\ \bibinfo {pages} {1169}
  (\bibinfo {year} {2013})}\BibitemShut {NoStop}%
\bibitem [{\citenamefont {Barends}\ \emph {et~al.}(2014)\citenamefont
  {Barends}, \citenamefont {Kelly}, \citenamefont {Megrant}, \citenamefont
  {Veitia}, \citenamefont {Sank}, \citenamefont {Jeffrey}, \citenamefont
  {White}, \citenamefont {Mutus}, \citenamefont {Fowler}, \citenamefont
  {Campbell}, \citenamefont {Chen}, \citenamefont {Chen}, \citenamefont
  {Chiaro}, \citenamefont {Dunsworth}, \citenamefont {Neill}, \citenamefont
  {O'Malley}, \citenamefont {Roushan}, \citenamefont {Vainsencher},
  \citenamefont {Wenner}, \citenamefont {Korotkov}, \citenamefont {Cleland},\
  and\ \citenamefont {Martinis}}]{Barends:2014fu}%
  \BibitemOpen
  \bibfield  {author} {\bibinfo {author} {\bibfnamefont {R.}~\bibnamefont
  {Barends}}, \bibinfo {author} {\bibfnamefont {J.}~\bibnamefont {Kelly}},
  \bibinfo {author} {\bibfnamefont {A.}~\bibnamefont {Megrant}}, \bibinfo
  {author} {\bibfnamefont {A.}~\bibnamefont {Veitia}}, \bibinfo {author}
  {\bibfnamefont {D.}~\bibnamefont {Sank}}, \bibinfo {author} {\bibfnamefont
  {E.}~\bibnamefont {Jeffrey}}, \bibinfo {author} {\bibfnamefont {T.~C.}\
  \bibnamefont {White}}, \bibinfo {author} {\bibfnamefont {J.}~\bibnamefont
  {Mutus}}, \bibinfo {author} {\bibfnamefont {A.~G.}\ \bibnamefont {Fowler}},
  \bibinfo {author} {\bibfnamefont {B.}~\bibnamefont {Campbell}}, \bibinfo
  {author} {\bibfnamefont {Y.}~\bibnamefont {Chen}}, \bibinfo {author}
  {\bibfnamefont {Z.}~\bibnamefont {Chen}}, \bibinfo {author} {\bibfnamefont
  {B.}~\bibnamefont {Chiaro}}, \bibinfo {author} {\bibfnamefont
  {A.}~\bibnamefont {Dunsworth}}, \bibinfo {author} {\bibfnamefont
  {C.}~\bibnamefont {Neill}}, \bibinfo {author} {\bibfnamefont
  {P.}~\bibnamefont {O'Malley}}, \bibinfo {author} {\bibfnamefont
  {P.}~\bibnamefont {Roushan}}, \bibinfo {author} {\bibfnamefont
  {A.}~\bibnamefont {Vainsencher}}, \bibinfo {author} {\bibfnamefont
  {J.}~\bibnamefont {Wenner}}, \bibinfo {author} {\bibfnamefont {A.~N.}\
  \bibnamefont {Korotkov}}, \bibinfo {author} {\bibfnamefont {A.~N.}\
  \bibnamefont {Cleland}}, \ and\ \bibinfo {author} {\bibfnamefont {J.~M.}\
  \bibnamefont {Martinis}},\ }\href@noop {} {\bibfield  {journal} {\bibinfo
  {journal} {Nature}\ }\textbf {\bibinfo {volume} {508}},\ \bibinfo {pages}
  {500} (\bibinfo {year} {2014})}\BibitemShut {NoStop}%
\bibitem [{\citenamefont {Bylander}\ \emph {et~al.}(2011)\citenamefont
  {Bylander}, \citenamefont {Gustavsson}, \citenamefont {Yan}, \citenamefont
  {Yoshihara}, \citenamefont {Harrabi}, \citenamefont {Fitch}, \citenamefont
  {Cory}, \citenamefont {Nakamura}, \citenamefont {Tsai},\ and\ \citenamefont
  {Oliver}}]{Bylander:2011wc}%
  \BibitemOpen
  \bibfield  {author} {\bibinfo {author} {\bibfnamefont {J.}~\bibnamefont
  {Bylander}}, \bibinfo {author} {\bibfnamefont {S.}~\bibnamefont
  {Gustavsson}}, \bibinfo {author} {\bibfnamefont {F.}~\bibnamefont {Yan}},
  \bibinfo {author} {\bibfnamefont {F.}~\bibnamefont {Yoshihara}}, \bibinfo
  {author} {\bibfnamefont {K.}~\bibnamefont {Harrabi}}, \bibinfo {author}
  {\bibfnamefont {G.}~\bibnamefont {Fitch}}, \bibinfo {author} {\bibfnamefont
  {D.~G.}\ \bibnamefont {Cory}}, \bibinfo {author} {\bibfnamefont
  {Y.}~\bibnamefont {Nakamura}}, \bibinfo {author} {\bibfnamefont {J.-S.}\
  \bibnamefont {Tsai}}, \ and\ \bibinfo {author} {\bibfnamefont {W.~D.}\
  \bibnamefont {Oliver}},\ }\href@noop {} {\bibfield  {journal} {\bibinfo
  {journal} {Nature Physics}\ }\textbf {\bibinfo {volume} {7}},\ \bibinfo
  {pages} {565} (\bibinfo {year} {2011})}\BibitemShut {NoStop}%
\bibitem [{\citenamefont {Kim}\ \emph {et~al.}(2011)\citenamefont {Kim},
  \citenamefont {Suri}, \citenamefont {Zaretskey}, \citenamefont {Novikov},
  \citenamefont {Osborn}, \citenamefont {Mizel}, \citenamefont {Wellstood},\
  and\ \citenamefont {Palmer}}]{Kim:2011uj}%
  \BibitemOpen
  \bibfield  {author} {\bibinfo {author} {\bibfnamefont {Z.}~\bibnamefont
  {Kim}}, \bibinfo {author} {\bibfnamefont {B.}~\bibnamefont {Suri}}, \bibinfo
  {author} {\bibfnamefont {V.}~\bibnamefont {Zaretskey}}, \bibinfo {author}
  {\bibfnamefont {S.}~\bibnamefont {Novikov}}, \bibinfo {author} {\bibfnamefont
  {K.~D.}\ \bibnamefont {Osborn}}, \bibinfo {author} {\bibfnamefont
  {A.}~\bibnamefont {Mizel}}, \bibinfo {author} {\bibfnamefont {F.~C.}\
  \bibnamefont {Wellstood}}, \ and\ \bibinfo {author} {\bibfnamefont {B.~S.}\
  \bibnamefont {Palmer}},\ }\href@noop {} {\bibfield  {journal} {\bibinfo
  {journal} {Physical Review Letters}\ }\textbf {\bibinfo {volume} {106}},\
  \bibinfo {pages} {120501} (\bibinfo {year} {2011})}\BibitemShut {NoStop}%
\bibitem [{\citenamefont {Manucharyan}\ \emph {et~al.}(2012)\citenamefont
  {Manucharyan}, \citenamefont {Masluk}, \citenamefont {Kamal}, \citenamefont
  {Koch}, \citenamefont {Glazman},\ and\ \citenamefont
  {Devoret}}]{Manucharyan:2012wm}%
  \BibitemOpen
  \bibfield  {author} {\bibinfo {author} {\bibfnamefont {V.~E.}\ \bibnamefont
  {Manucharyan}}, \bibinfo {author} {\bibfnamefont {N.~A.}\ \bibnamefont
  {Masluk}}, \bibinfo {author} {\bibfnamefont {A.}~\bibnamefont {Kamal}},
  \bibinfo {author} {\bibfnamefont {J.}~\bibnamefont {Koch}}, \bibinfo {author}
  {\bibfnamefont {L.~I.}\ \bibnamefont {Glazman}}, \ and\ \bibinfo {author}
  {\bibfnamefont {M.~H.}\ \bibnamefont {Devoret}},\ }\href@noop {} {\bibfield
  {journal} {\bibinfo  {journal} {Physical Review B}\ }\textbf {\bibinfo
  {volume} {85}},\ \bibinfo {pages} {024521} (\bibinfo {year}
  {2012})}\BibitemShut {NoStop}%
\bibitem [{\citenamefont {Paik}\ \emph {et~al.}(2011)\citenamefont {Paik},
  \citenamefont {Schuster}, \citenamefont {Bishop}, \citenamefont {Kirchmair},
  \citenamefont {Catelani}, \citenamefont {Sears}, \citenamefont {Johnson},
  \citenamefont {Reagor}, \citenamefont {Frunzio}, \citenamefont {Glazman},
  \citenamefont {Girvin}, \citenamefont {Devoret},\ and\ \citenamefont
  {Schoelkopf}}]{Paik:2011hd}%
  \BibitemOpen
  \bibfield  {author} {\bibinfo {author} {\bibfnamefont {H.}~\bibnamefont
  {Paik}}, \bibinfo {author} {\bibfnamefont {D.~I.}\ \bibnamefont {Schuster}},
  \bibinfo {author} {\bibfnamefont {L.~S.}\ \bibnamefont {Bishop}}, \bibinfo
  {author} {\bibfnamefont {G.}~\bibnamefont {Kirchmair}}, \bibinfo {author}
  {\bibfnamefont {G.}~\bibnamefont {Catelani}}, \bibinfo {author}
  {\bibfnamefont {A.~P.}\ \bibnamefont {Sears}}, \bibinfo {author}
  {\bibfnamefont {B.~R.}\ \bibnamefont {Johnson}}, \bibinfo {author}
  {\bibfnamefont {M.~J.}\ \bibnamefont {Reagor}}, \bibinfo {author}
  {\bibfnamefont {L.}~\bibnamefont {Frunzio}}, \bibinfo {author} {\bibfnamefont
  {L.~I.}\ \bibnamefont {Glazman}}, \bibinfo {author} {\bibfnamefont {S.~M.}\
  \bibnamefont {Girvin}}, \bibinfo {author} {\bibfnamefont {M.~H.}\
  \bibnamefont {Devoret}}, \ and\ \bibinfo {author} {\bibfnamefont {R.~J.}\
  \bibnamefont {Schoelkopf}},\ }\href@noop {} {\bibfield  {journal} {\bibinfo
  {journal} {Physical Review Letters}\ }\textbf {\bibinfo {volume} {107}},\
  \bibinfo {pages} {240501} (\bibinfo {year} {2011})}\BibitemShut {NoStop}%
\bibitem [{\citenamefont {Doh}\ \emph {et~al.}(2005)\citenamefont {Doh},
  \citenamefont {van Dam}, \citenamefont {Roest}, \citenamefont {Bakkers},
  \citenamefont {Kouwenhoven},\ and\ \citenamefont
  {De~Franceschi}}]{Doh:878391}%
  \BibitemOpen
  \bibfield  {author} {\bibinfo {author} {\bibfnamefont {Y.~J.}\ \bibnamefont
  {Doh}}, \bibinfo {author} {\bibfnamefont {J.~A.}\ \bibnamefont {van Dam}},
  \bibinfo {author} {\bibfnamefont {A.~L.}\ \bibnamefont {Roest}}, \bibinfo
  {author} {\bibfnamefont {E.~P. A.~M.}\ \bibnamefont {Bakkers}}, \bibinfo
  {author} {\bibfnamefont {L.~P.}\ \bibnamefont {Kouwenhoven}}, \ and\ \bibinfo
  {author} {\bibfnamefont {S.}~\bibnamefont {De~Franceschi}},\ }\href@noop {}
  {\bibfield  {journal} {\bibinfo  {journal} {Science}\ }\textbf {\bibinfo
  {volume} {309}},\ \bibinfo {pages} {272} (\bibinfo {year}
  {2005})}\BibitemShut {NoStop}%
\bibitem [{\citenamefont {Abay}\ \emph {et~al.}(2014)\citenamefont {Abay},
  \citenamefont {Persson}, \citenamefont {Nilsson}, \citenamefont {Wu},
  \citenamefont {Xu}, \citenamefont {Fogelstr{\"o}m}, \citenamefont
  {Shumeiko},\ and\ \citenamefont {Delsing}}]{Abay:2014ed}%
  \BibitemOpen
  \bibfield  {author} {\bibinfo {author} {\bibfnamefont {S.}~\bibnamefont
  {Abay}}, \bibinfo {author} {\bibfnamefont {D.}~\bibnamefont {Persson}},
  \bibinfo {author} {\bibfnamefont {H.}~\bibnamefont {Nilsson}}, \bibinfo
  {author} {\bibfnamefont {F.}~\bibnamefont {Wu}}, \bibinfo {author}
  {\bibfnamefont {H.~Q.}\ \bibnamefont {Xu}}, \bibinfo {author} {\bibfnamefont
  {M.}~\bibnamefont {Fogelstr{\"o}m}}, \bibinfo {author} {\bibfnamefont
  {V.}~\bibnamefont {Shumeiko}}, \ and\ \bibinfo {author} {\bibfnamefont
  {P.}~\bibnamefont {Delsing}},\ }\href@noop {} {\bibfield  {journal} {\bibinfo
   {journal} {Physical Review B}\ }\textbf {\bibinfo {volume} {89}},\ \bibinfo
  {pages} {214508} (\bibinfo {year} {2014})}\BibitemShut {NoStop}%
\bibitem [{\citenamefont {Krogstrup}\ \emph {et~al.}()\citenamefont
  {Krogstrup}, \citenamefont {Ziino}, \citenamefont {Albrecht}, \citenamefont
  {Madsen}, \citenamefont {Johnson}, \citenamefont {Nyg{\aa}rd}, \citenamefont
  {Marcus},\ and\ \citenamefont {Jespersen}}]{Krogstrup:2014va}%
  \BibitemOpen
  \bibfield  {author} {\bibinfo {author} {\bibfnamefont {P.}~\bibnamefont
  {Krogstrup}}, \bibinfo {author} {\bibfnamefont {N.~L.~B.}\ \bibnamefont
  {Ziino}}, \bibinfo {author} {\bibfnamefont {S.~M.}\ \bibnamefont {Albrecht}},
  \bibinfo {author} {\bibfnamefont {M.~H.}\ \bibnamefont {Madsen}}, \bibinfo
  {author} {\bibfnamefont {E.}~\bibnamefont {Johnson}}, \bibinfo {author}
  {\bibfnamefont {J.}~\bibnamefont {Nyg{\aa}rd}}, \bibinfo {author}
  {\bibfnamefont {C.~M.}\ \bibnamefont {Marcus}}, \ and\ \bibinfo {author}
  {\bibfnamefont {T.~S.}\ \bibnamefont {Jespersen}},\ }\href@noop {} {\bibinfo
  {journal} {Nature Materials, Advance Online Publication}\ }\BibitemShut
  {NoStop}%
\bibitem [{\citenamefont {Chang}\ \emph {et~al.}(2015)\citenamefont {Chang},
  \citenamefont {Albrecht}, \citenamefont {Jespersen}, \citenamefont
  {Kuemmeth}, \citenamefont {Krogstrup}, \citenamefont {Nyg{\aa}rd},\ and\
  \citenamefont {Marcus}}]{Chang:2014vf}%
  \BibitemOpen
\bibfield  {journal} {  }\bibfield  {author} {\bibinfo {author} {\bibfnamefont
  {W.}~\bibnamefont {Chang}}, \bibinfo {author} {\bibfnamefont {S.~M.}\
  \bibnamefont {Albrecht}}, \bibinfo {author} {\bibfnamefont {T.~S.}\
  \bibnamefont {Jespersen}}, \bibinfo {author} {\bibfnamefont {F.}~\bibnamefont
  {Kuemmeth}}, \bibinfo {author} {\bibfnamefont {P.}~\bibnamefont {Krogstrup}},
  \bibinfo {author} {\bibfnamefont {J.}~\bibnamefont {Nyg{\aa}rd}}, \ and\
  \bibinfo {author} {\bibfnamefont {C.~M.}\ \bibnamefont {Marcus}},\
  }\href@noop {} {\bibfield  {journal} {\bibinfo  {journal} {Nature
  Nanotechnology}\ }\textbf {\bibinfo {volume} {10}},\ \bibinfo {pages} {232}
  (\bibinfo {year} {2015})}\BibitemShut {NoStop}%
\bibitem [{\citenamefont {Shim}\ and\ \citenamefont {Tahan}(2014)}]{Shim:2014}%
  \BibitemOpen
  \bibfield  {author} {\bibinfo {author} {\bibfnamefont {Y.-P.}\ \bibnamefont
  {Shim}}\ and\ \bibinfo {author} {\bibfnamefont {C.}~\bibnamefont {Tahan}},\
  }\href {http://dx.doi.org/10.1038/ncomms5225} {\bibfield  {journal} {\bibinfo
   {journal} {Nature Communications}\ }\textbf {\bibinfo {volume} {5}},\
  (\bibinfo {year} {2014})}\BibitemShut {NoStop}%
\bibitem [{\citenamefont {Koch}\ \emph {et~al.}(2007)\citenamefont {Koch},
  \citenamefont {Yu}, \citenamefont {Gambetta}, \citenamefont {Houck},
  \citenamefont {Schuster}, \citenamefont {Majer}, \citenamefont {Blais},
  \citenamefont {Devoret}, \citenamefont {Girvin},\ and\ \citenamefont
  {Schoelkopf}}]{Koch:2007gz}%
  \BibitemOpen
  \bibfield  {author} {\bibinfo {author} {\bibfnamefont {J.}~\bibnamefont
  {Koch}}, \bibinfo {author} {\bibfnamefont {T.}~\bibnamefont {Yu}}, \bibinfo
  {author} {\bibfnamefont {J.}~\bibnamefont {Gambetta}}, \bibinfo {author}
  {\bibfnamefont {A.}~\bibnamefont {Houck}}, \bibinfo {author} {\bibfnamefont
  {D.}~\bibnamefont {Schuster}}, \bibinfo {author} {\bibfnamefont
  {J.}~\bibnamefont {Majer}}, \bibinfo {author} {\bibfnamefont
  {A.}~\bibnamefont {Blais}}, \bibinfo {author} {\bibfnamefont
  {M.}~\bibnamefont {Devoret}}, \bibinfo {author} {\bibfnamefont
  {S.}~\bibnamefont {Girvin}}, \ and\ \bibinfo {author} {\bibfnamefont
  {R.}~\bibnamefont {Schoelkopf}},\ }\href@noop {} {\bibfield  {journal}
  {\bibinfo  {journal} {Physical Review A}\ }\textbf {\bibinfo {volume} {76}},\
  \bibinfo {pages} {042319} (\bibinfo {year} {2007})}\BibitemShut {NoStop}%
\bibitem [{\citenamefont {Houck}\ \emph {et~al.}(2007)\citenamefont {Houck},
  \citenamefont {Schuster}, \citenamefont {Gambetta}, \citenamefont {Schreier},
  \citenamefont {Johnson}, \citenamefont {Chow}, \citenamefont {Frunzio},
  \citenamefont {Majer}, \citenamefont {Devoret}, \citenamefont {Girvin},\ and\
  \citenamefont {Schoelkopf}}]{Houck:2007dy}%
  \BibitemOpen
  \bibfield  {author} {\bibinfo {author} {\bibfnamefont {A.~A.}\ \bibnamefont
  {Houck}}, \bibinfo {author} {\bibfnamefont {D.~I.}\ \bibnamefont {Schuster}},
  \bibinfo {author} {\bibfnamefont {J.~M.}\ \bibnamefont {Gambetta}}, \bibinfo
  {author} {\bibfnamefont {J.~A.}\ \bibnamefont {Schreier}}, \bibinfo {author}
  {\bibfnamefont {B.~R.}\ \bibnamefont {Johnson}}, \bibinfo {author}
  {\bibfnamefont {J.~M.}\ \bibnamefont {Chow}}, \bibinfo {author}
  {\bibfnamefont {L.}~\bibnamefont {Frunzio}}, \bibinfo {author} {\bibfnamefont
  {J.}~\bibnamefont {Majer}}, \bibinfo {author} {\bibfnamefont {M.~H.}\
  \bibnamefont {Devoret}}, \bibinfo {author} {\bibfnamefont {S.~M.}\
  \bibnamefont {Girvin}}, \ and\ \bibinfo {author} {\bibfnamefont {R.~J.}\
  \bibnamefont {Schoelkopf}},\ }\href@noop {} {\bibfield  {journal} {\bibinfo
  {journal} {Nature}\ }\textbf {\bibinfo {volume} {449}},\ \bibinfo {pages}
  {328} (\bibinfo {year} {2007})}\BibitemShut {NoStop}%
\bibitem [{\citenamefont {de~Lange}\ \emph {et~al.}()\citenamefont {de~Lange},
  \citenamefont {van Heck}, \citenamefont {Brunol}, \citenamefont {van
  Woerkom}, \citenamefont {Geresdil}, \citenamefont {Plissard}, \citenamefont
  {Bakkers}, \citenamefont {Akhmerov},\ and\ \citenamefont
  {DiCarlo}}]{deLange:2014}%
  \BibitemOpen
  \bibfield  {author} {\bibinfo {author} {\bibfnamefont {G.}~\bibnamefont
  {de~Lange}}, \bibinfo {author} {\bibfnamefont {B.}~\bibnamefont {van Heck}},
  \bibinfo {author} {\bibfnamefont {A.}~\bibnamefont {Brunol}}, \bibinfo
  {author} {\bibfnamefont {D.}~\bibnamefont {van Woerkom}}, \bibinfo {author}
  {\bibfnamefont {A.}~\bibnamefont {Geresdil}}, \bibinfo {author}
  {\bibfnamefont {S.~R.}\ \bibnamefont {Plissard}}, \bibinfo {author}
  {\bibfnamefont {E.~P. A.~M.}\ \bibnamefont {Bakkers}}, \bibinfo {author}
  {\bibfnamefont {A.~R.}\ \bibnamefont {Akhmerov}}, \ and\ \bibinfo {author}
  {\bibfnamefont {L.}~\bibnamefont {DiCarlo}},\ }\href@noop {} {\bibinfo
  {journal} {\textit{submitted}}\ }\BibitemShut {NoStop}%
\bibitem [{\citenamefont {Barends}\ \emph {et~al.}(2013)\citenamefont
  {Barends}, \citenamefont {Kelly}, \citenamefont {Megrant}, \citenamefont
  {Sank}, \citenamefont {Jeffrey}, \citenamefont {Chen}, \citenamefont {Yin},
  \citenamefont {Chiaro}, \citenamefont {Mutus}, \citenamefont {Neill},
  \citenamefont {O'Malley}, \citenamefont {Roushan}, \citenamefont {Wenner},
  \citenamefont {White}, \citenamefont {Cleland},\ and\ \citenamefont
  {Martinis}}]{Barends:2013kz}%
  \BibitemOpen
\bibfield  {journal} {  }\bibfield  {author} {\bibinfo {author} {\bibfnamefont
  {R.}~\bibnamefont {Barends}}, \bibinfo {author} {\bibfnamefont
  {J.}~\bibnamefont {Kelly}}, \bibinfo {author} {\bibfnamefont
  {A.}~\bibnamefont {Megrant}}, \bibinfo {author} {\bibfnamefont
  {D.}~\bibnamefont {Sank}}, \bibinfo {author} {\bibfnamefont {E.}~\bibnamefont
  {Jeffrey}}, \bibinfo {author} {\bibfnamefont {Y.}~\bibnamefont {Chen}},
  \bibinfo {author} {\bibfnamefont {Y.}~\bibnamefont {Yin}}, \bibinfo {author}
  {\bibfnamefont {B.}~\bibnamefont {Chiaro}}, \bibinfo {author} {\bibfnamefont
  {J.}~\bibnamefont {Mutus}}, \bibinfo {author} {\bibfnamefont
  {C.}~\bibnamefont {Neill}}, \bibinfo {author} {\bibfnamefont
  {P.}~\bibnamefont {O'Malley}}, \bibinfo {author} {\bibfnamefont
  {P.}~\bibnamefont {Roushan}}, \bibinfo {author} {\bibfnamefont
  {J.}~\bibnamefont {Wenner}}, \bibinfo {author} {\bibfnamefont {T.~C.}\
  \bibnamefont {White}}, \bibinfo {author} {\bibfnamefont {A.~N.}\ \bibnamefont
  {Cleland}}, \ and\ \bibinfo {author} {\bibfnamefont {J.~M.}\ \bibnamefont
  {Martinis}},\ }\href@noop {} {\bibfield  {journal} {\bibinfo  {journal}
  {Physical Review Letters}\ }\textbf {\bibinfo {volume} {111}},\ \bibinfo
  {pages} {080502} (\bibinfo {year} {2013})}\BibitemShut {NoStop}%
\bibitem [{Hom()}]{Homodyne}%
  \BibitemOpen
  \href@noop {} {}\bibinfo {note} {The data in Fig.\ 2 were acquired using a
  vector network analyzer. For the Sample 1 data in Figs.\ 3 and 4 we mix down
  to dc and sample the homodyne response, $V_{\mathrm{H}}$. For the Sample 2
  data in Fig.\ 4 we mix down to an intermediate frequency before sampling and
  then perform digital homodyne to extract the cavity phase
  response.}\BibitemShut {Stop}%
\bibitem [{\citenamefont {Wallraff}\ \emph {et~al.}(2004)\citenamefont
  {Wallraff}, \citenamefont {Schuster}, \citenamefont {Blais}, \citenamefont
  {Frunzio}, \citenamefont {Huang}, \citenamefont {Majer}, \citenamefont
  {Kumar}, \citenamefont {Girvin},\ and\ \citenamefont
  {Schoelkopf}}]{Wallraff:2004vt}%
  \BibitemOpen
  \bibfield  {author} {\bibinfo {author} {\bibfnamefont {A.}~\bibnamefont
  {Wallraff}}, \bibinfo {author} {\bibfnamefont {D.~I.}\ \bibnamefont
  {Schuster}}, \bibinfo {author} {\bibfnamefont {A.}~\bibnamefont {Blais}},
  \bibinfo {author} {\bibfnamefont {L.}~\bibnamefont {Frunzio}}, \bibinfo
  {author} {\bibfnamefont {R.-S.}\ \bibnamefont {Huang}}, \bibinfo {author}
  {\bibfnamefont {J.}~\bibnamefont {Majer}}, \bibinfo {author} {\bibfnamefont
  {S.}~\bibnamefont {Kumar}}, \bibinfo {author} {\bibfnamefont {S.~M.}\
  \bibnamefont {Girvin}}, \ and\ \bibinfo {author} {\bibfnamefont {R.~J.}\
  \bibnamefont {Schoelkopf}},\ }\href@noop {} {\bibfield  {journal} {\bibinfo
  {journal} {Nature}\ }\textbf {\bibinfo {volume} {431}},\ \bibinfo {pages}
  {162} (\bibinfo {year} {2004})}\BibitemShut {NoStop}%
\bibitem [{\citenamefont {DiCarlo}\ \emph {et~al.}(2009)\citenamefont
  {DiCarlo}, \citenamefont {Chow}, \citenamefont {Gambetta}, \citenamefont
  {Bishop}, \citenamefont {Johnson}, \citenamefont {Schuster}, \citenamefont
  {Majer}, \citenamefont {Blais}, \citenamefont {Frunzio}, \citenamefont
  {Girvin},\ and\ \citenamefont {Schoelkopf}}]{DiCarlo:2009ja}%
  \BibitemOpen
  \bibfield  {author} {\bibinfo {author} {\bibfnamefont {L.}~\bibnamefont
  {DiCarlo}}, \bibinfo {author} {\bibfnamefont {J.~M.}\ \bibnamefont {Chow}},
  \bibinfo {author} {\bibfnamefont {J.~M.}\ \bibnamefont {Gambetta}}, \bibinfo
  {author} {\bibfnamefont {L.~S.}\ \bibnamefont {Bishop}}, \bibinfo {author}
  {\bibfnamefont {B.~R.}\ \bibnamefont {Johnson}}, \bibinfo {author}
  {\bibfnamefont {D.~I.}\ \bibnamefont {Schuster}}, \bibinfo {author}
  {\bibfnamefont {J.}~\bibnamefont {Majer}}, \bibinfo {author} {\bibfnamefont
  {A.}~\bibnamefont {Blais}}, \bibinfo {author} {\bibfnamefont
  {L.}~\bibnamefont {Frunzio}}, \bibinfo {author} {\bibfnamefont {S.~M.}\
  \bibnamefont {Girvin}}, \ and\ \bibinfo {author} {\bibfnamefont {R.~J.}\
  \bibnamefont {Schoelkopf}},\ }\href@noop {} {\bibfield  {journal} {\bibinfo
  {journal} {Nature}\ }\textbf {\bibinfo {volume} {460}},\ \bibinfo {pages}
  {240} (\bibinfo {year} {2009})}\BibitemShut {NoStop}%
\bibitem [{\citenamefont {Houck}\ \emph {et~al.}(2008)\citenamefont {Houck},
  \citenamefont {Schreier}, \citenamefont {Johnson}, \citenamefont {Chow},
  \citenamefont {Koch}, \citenamefont {Gambetta}, \citenamefont {Schuster},
  \citenamefont {Frunzio}, \citenamefont {Devoret}, \citenamefont {Girvin},\
  and\ \citenamefont {Schoelkopf}}]{Houck:2008je}%
  \BibitemOpen
  \bibfield  {author} {\bibinfo {author} {\bibfnamefont {A.~A.}\ \bibnamefont
  {Houck}}, \bibinfo {author} {\bibfnamefont {J.~A.}\ \bibnamefont {Schreier}},
  \bibinfo {author} {\bibfnamefont {B.~R.}\ \bibnamefont {Johnson}}, \bibinfo
  {author} {\bibfnamefont {J.~M.}\ \bibnamefont {Chow}}, \bibinfo {author}
  {\bibfnamefont {J.}~\bibnamefont {Koch}}, \bibinfo {author} {\bibfnamefont
  {J.~M.}\ \bibnamefont {Gambetta}}, \bibinfo {author} {\bibfnamefont {D.~I.}\
  \bibnamefont {Schuster}}, \bibinfo {author} {\bibfnamefont {L.}~\bibnamefont
  {Frunzio}}, \bibinfo {author} {\bibfnamefont {M.~H.}\ \bibnamefont
  {Devoret}}, \bibinfo {author} {\bibfnamefont {S.~M.}\ \bibnamefont {Girvin}},
  \ and\ \bibinfo {author} {\bibfnamefont {R.~J.}\ \bibnamefont {Schoelkopf}},\
  }\href@noop {} {\bibfield  {journal} {\bibinfo  {journal} {Physical Review
  Letters}\ }\textbf {\bibinfo {volume} {101}},\ \bibinfo {pages} {080502}
  (\bibinfo {year} {2008})}\BibitemShut {NoStop}%
\bibitem [{\citenamefont {Chang}\ \emph {et~al.}(2013)\citenamefont {Chang},
  \citenamefont {Vissers}, \citenamefont {C{\'o}rcoles}, \citenamefont
  {Sandberg}, \citenamefont {Gao}, \citenamefont {Abraham}, \citenamefont
  {Chow}, \citenamefont {Gambetta}, \citenamefont {Rothwell}, \citenamefont
  {Keefe}, \citenamefont {Steffen},\ and\ \citenamefont
  {Pappas}}]{Chang:2013dw}%
  \BibitemOpen
  \bibfield  {author} {\bibinfo {author} {\bibfnamefont {J.~B.}\ \bibnamefont
  {Chang}}, \bibinfo {author} {\bibfnamefont {M.~R.}\ \bibnamefont {Vissers}},
  \bibinfo {author} {\bibfnamefont {A.~D.}\ \bibnamefont {C{\'o}rcoles}},
  \bibinfo {author} {\bibfnamefont {M.}~\bibnamefont {Sandberg}}, \bibinfo
  {author} {\bibfnamefont {J.}~\bibnamefont {Gao}}, \bibinfo {author}
  {\bibfnamefont {D.~W.}\ \bibnamefont {Abraham}}, \bibinfo {author}
  {\bibfnamefont {J.~M.}\ \bibnamefont {Chow}}, \bibinfo {author}
  {\bibfnamefont {J.~M.}\ \bibnamefont {Gambetta}}, \bibinfo {author}
  {\bibfnamefont {M.~B.}\ \bibnamefont {Rothwell}}, \bibinfo {author}
  {\bibfnamefont {G.~A.}\ \bibnamefont {Keefe}}, \bibinfo {author}
  {\bibfnamefont {M.}~\bibnamefont {Steffen}}, \ and\ \bibinfo {author}
  {\bibfnamefont {D.~P.}\ \bibnamefont {Pappas}},\ }\href@noop {} {\bibfield
  {journal} {\bibinfo  {journal} {Applied Physics Letters}\ }\textbf {\bibinfo
  {volume} {103}},\ \bibinfo {pages} {012602} (\bibinfo {year}
  {2013})}\BibitemShut {NoStop}%
\bibitem [{\citenamefont {O'Connell}\ \emph {et~al.}(2008)\citenamefont
  {O'Connell}, \citenamefont {Ansmann}, \citenamefont {Bialczak}, \citenamefont
  {Hofheinz}, \citenamefont {Katz}, \citenamefont {Lucero}, \citenamefont
  {McKenney}, \citenamefont {Neeley}, \citenamefont {Wang}, \citenamefont
  {Weig}, \citenamefont {Cleland},\ and\ \citenamefont
  {Martinis}}]{OConnell:2008jt}%
  \BibitemOpen
  \bibfield  {author} {\bibinfo {author} {\bibfnamefont {A.~D.}\ \bibnamefont
  {O'Connell}}, \bibinfo {author} {\bibfnamefont {M.}~\bibnamefont {Ansmann}},
  \bibinfo {author} {\bibfnamefont {R.~C.}\ \bibnamefont {Bialczak}}, \bibinfo
  {author} {\bibfnamefont {M.}~\bibnamefont {Hofheinz}}, \bibinfo {author}
  {\bibfnamefont {N.}~\bibnamefont {Katz}}, \bibinfo {author} {\bibfnamefont
  {E.}~\bibnamefont {Lucero}}, \bibinfo {author} {\bibfnamefont
  {C.}~\bibnamefont {McKenney}}, \bibinfo {author} {\bibfnamefont
  {M.}~\bibnamefont {Neeley}}, \bibinfo {author} {\bibfnamefont
  {H.}~\bibnamefont {Wang}}, \bibinfo {author} {\bibfnamefont {E.~M.}\
  \bibnamefont {Weig}}, \bibinfo {author} {\bibfnamefont {A.~N.}\ \bibnamefont
  {Cleland}}, \ and\ \bibinfo {author} {\bibfnamefont {J.~M.}\ \bibnamefont
  {Martinis}},\ }\href@noop {} {\bibfield  {journal} {\bibinfo  {journal}
  {Applied Physics Letters}\ }\textbf {\bibinfo {volume} {92}},\ \bibinfo
  {pages} {112903} (\bibinfo {year} {2008})}\BibitemShut {NoStop}%
\bibitem [{\citenamefont {Quintana}\ \emph {et~al.}(2014)\citenamefont
  {Quintana}, \citenamefont {Megrant}, \citenamefont {Chen}, \citenamefont
  {Dunsworth}, \citenamefont {Chiaro}, \citenamefont {Barends}, \citenamefont
  {Campbell}, \citenamefont {Chen}, \citenamefont {Hoi}, \citenamefont
  {Jeffrey}, \citenamefont {Kelly}, \citenamefont {Mutus}, \citenamefont
  {O'Malley}, \citenamefont {Neill}, \citenamefont {Roushan}, \citenamefont
  {Sank}, \citenamefont {Vainsencher}, \citenamefont {Wenner}, \citenamefont
  {White}, \citenamefont {Cleland},\ and\ \citenamefont
  {Martinis}}]{Quintana:2014jp}%
  \BibitemOpen
  \bibfield  {author} {\bibinfo {author} {\bibfnamefont {C.~M.}\ \bibnamefont
  {Quintana}}, \bibinfo {author} {\bibfnamefont {A.}~\bibnamefont {Megrant}},
  \bibinfo {author} {\bibfnamefont {Z.}~\bibnamefont {Chen}}, \bibinfo {author}
  {\bibfnamefont {A.}~\bibnamefont {Dunsworth}}, \bibinfo {author}
  {\bibfnamefont {B.}~\bibnamefont {Chiaro}}, \bibinfo {author} {\bibfnamefont
  {R.}~\bibnamefont {Barends}}, \bibinfo {author} {\bibfnamefont
  {B.}~\bibnamefont {Campbell}}, \bibinfo {author} {\bibfnamefont
  {Y.}~\bibnamefont {Chen}}, \bibinfo {author} {\bibfnamefont {I.~C.}\
  \bibnamefont {Hoi}}, \bibinfo {author} {\bibfnamefont {E.}~\bibnamefont
  {Jeffrey}}, \bibinfo {author} {\bibfnamefont {J.}~\bibnamefont {Kelly}},
  \bibinfo {author} {\bibfnamefont {J.~Y.}\ \bibnamefont {Mutus}}, \bibinfo
  {author} {\bibfnamefont {P.~J.~J.}\ \bibnamefont {O'Malley}}, \bibinfo
  {author} {\bibfnamefont {C.}~\bibnamefont {Neill}}, \bibinfo {author}
  {\bibfnamefont {P.}~\bibnamefont {Roushan}}, \bibinfo {author} {\bibfnamefont
  {D.}~\bibnamefont {Sank}}, \bibinfo {author} {\bibfnamefont {A.}~\bibnamefont
  {Vainsencher}}, \bibinfo {author} {\bibfnamefont {J.}~\bibnamefont {Wenner}},
  \bibinfo {author} {\bibfnamefont {T.~C.}\ \bibnamefont {White}}, \bibinfo
  {author} {\bibfnamefont {A.~N.}\ \bibnamefont {Cleland}}, \ and\ \bibinfo
  {author} {\bibfnamefont {J.~M.}\ \bibnamefont {Martinis}},\ }\href@noop {}
  {\bibfield  {journal} {\bibinfo  {journal} {Applied Physics Letters}\
  }\textbf {\bibinfo {volume} {105}},\ \bibinfo {pages} {062601} (\bibinfo
  {year} {2014})}\BibitemShut {NoStop}%
\bibitem [{\citenamefont {C{\'o}rcoles}\ \emph {et~al.}(2011)\citenamefont
  {C{\'o}rcoles}, \citenamefont {Chow}, \citenamefont {Gambetta}, \citenamefont
  {Rigetti}, \citenamefont {Rozen}, \citenamefont {Keefe}, \citenamefont
  {Rothwell}, \citenamefont {Ketchen},\ and\ \citenamefont
  {Steffen}}]{Corcoles:2011is}%
  \BibitemOpen
  \bibfield  {author} {\bibinfo {author} {\bibfnamefont {A.~D.}\ \bibnamefont
  {C{\'o}rcoles}}, \bibinfo {author} {\bibfnamefont {J.~M.}\ \bibnamefont
  {Chow}}, \bibinfo {author} {\bibfnamefont {J.~M.}\ \bibnamefont {Gambetta}},
  \bibinfo {author} {\bibfnamefont {C.}~\bibnamefont {Rigetti}}, \bibinfo
  {author} {\bibfnamefont {J.~R.}\ \bibnamefont {Rozen}}, \bibinfo {author}
  {\bibfnamefont {G.~A.}\ \bibnamefont {Keefe}}, \bibinfo {author}
  {\bibfnamefont {M.~B.}\ \bibnamefont {Rothwell}}, \bibinfo {author}
  {\bibfnamefont {M.~B.}\ \bibnamefont {Ketchen}}, \ and\ \bibinfo {author}
  {\bibfnamefont {M.}~\bibnamefont {Steffen}},\ }\href@noop {} {\bibfield
  {journal} {\bibinfo  {journal} {Applied Physics Letters}\ }\textbf {\bibinfo
  {volume} {99}},\ \bibinfo {pages} {181906} (\bibinfo {year}
  {2011})}\BibitemShut {NoStop}%
\bibitem [{\citenamefont {Barends}\ \emph {et~al.}(2011)\citenamefont
  {Barends}, \citenamefont {Wenner}, \citenamefont {Lenander}, \citenamefont
  {Chen}, \citenamefont {Bialczak}, \citenamefont {Kelly}, \citenamefont
  {Lucero}, \citenamefont {O'Malley}, \citenamefont {Mariantoni}, \citenamefont
  {Sank}, \citenamefont {Wang}, \citenamefont {White}, \citenamefont {Yin},
  \citenamefont {Zhao}, \citenamefont {Cleland}, \citenamefont {Martinis},\
  and\ \citenamefont {Baselmans}}]{Barends:2011eh}%
  \BibitemOpen
  \bibfield  {author} {\bibinfo {author} {\bibfnamefont {R.}~\bibnamefont
  {Barends}}, \bibinfo {author} {\bibfnamefont {J.}~\bibnamefont {Wenner}},
  \bibinfo {author} {\bibfnamefont {M.}~\bibnamefont {Lenander}}, \bibinfo
  {author} {\bibfnamefont {Y.}~\bibnamefont {Chen}}, \bibinfo {author}
  {\bibfnamefont {R.~C.}\ \bibnamefont {Bialczak}}, \bibinfo {author}
  {\bibfnamefont {J.}~\bibnamefont {Kelly}}, \bibinfo {author} {\bibfnamefont
  {E.}~\bibnamefont {Lucero}}, \bibinfo {author} {\bibfnamefont
  {P.}~\bibnamefont {O'Malley}}, \bibinfo {author} {\bibfnamefont
  {M.}~\bibnamefont {Mariantoni}}, \bibinfo {author} {\bibfnamefont
  {D.}~\bibnamefont {Sank}}, \bibinfo {author} {\bibfnamefont {H.}~\bibnamefont
  {Wang}}, \bibinfo {author} {\bibfnamefont {T.~C.}\ \bibnamefont {White}},
  \bibinfo {author} {\bibfnamefont {Y.}~\bibnamefont {Yin}}, \bibinfo {author}
  {\bibfnamefont {J.}~\bibnamefont {Zhao}}, \bibinfo {author} {\bibfnamefont
  {A.~N.}\ \bibnamefont {Cleland}}, \bibinfo {author} {\bibfnamefont {J.~M.}\
  \bibnamefont {Martinis}}, \ and\ \bibinfo {author} {\bibfnamefont {J.~J.~A.}\
  \bibnamefont {Baselmans}},\ }\href@noop {} {\bibfield  {journal} {\bibinfo
  {journal} {Applied Physics Letters}\ }\textbf {\bibinfo {volume} {99}},\
  \bibinfo {pages} {113507} (\bibinfo {year} {2011})}\BibitemShut {NoStop}%
\bibitem [{\citenamefont {Ford}\ \emph {et~al.}(2012)\citenamefont {Ford},
  \citenamefont {Kumar}, \citenamefont {Kapadia}, \citenamefont {Guo},\ and\
  \citenamefont {Javey}}]{Ford:2012cx}%
  \BibitemOpen
  \bibfield  {author} {\bibinfo {author} {\bibfnamefont {A.~C.}\ \bibnamefont
  {Ford}}, \bibinfo {author} {\bibfnamefont {S.~B.}\ \bibnamefont {Kumar}},
  \bibinfo {author} {\bibfnamefont {R.}~\bibnamefont {Kapadia}}, \bibinfo
  {author} {\bibfnamefont {J.}~\bibnamefont {Guo}}, \ and\ \bibinfo {author}
  {\bibfnamefont {A.}~\bibnamefont {Javey}},\ }\href@noop {} {\bibfield
  {journal} {\bibinfo  {journal} {Nano Letters}\ }\textbf {\bibinfo {volume}
  {12}},\ \bibinfo {pages} {1340} (\bibinfo {year} {2012})}\BibitemShut
  {NoStop}%
\bibitem [{\citenamefont {Ward}\ \emph {et~al.}(2013)\citenamefont {Ward},
  \citenamefont {Savage}, \citenamefont {Lagally}, \citenamefont
  {Coppersmith},\ and\ \citenamefont {Eriksson}}]{Ward:2013hq}%
  \BibitemOpen
  \bibfield  {author} {\bibinfo {author} {\bibfnamefont {D.~R.}\ \bibnamefont
  {Ward}}, \bibinfo {author} {\bibfnamefont {D.~E.}\ \bibnamefont {Savage}},
  \bibinfo {author} {\bibfnamefont {M.~G.}\ \bibnamefont {Lagally}}, \bibinfo
  {author} {\bibfnamefont {S.~N.}\ \bibnamefont {Coppersmith}}, \ and\ \bibinfo
  {author} {\bibfnamefont {M.~A.}\ \bibnamefont {Eriksson}},\ }\href@noop {}
  {\bibfield  {journal} {\bibinfo  {journal} {Applied Physics Letters}\
  }\textbf {\bibinfo {volume} {102}},\ \bibinfo {pages} {213107} (\bibinfo
  {year} {2013})}\BibitemShut {NoStop}%
\bibitem [{\citenamefont {Al-Taie}\ \emph {et~al.}(2013)\citenamefont
  {Al-Taie}, \citenamefont {Smith}, \citenamefont {Xu}, \citenamefont {See},
  \citenamefont {Griffiths}, \citenamefont {Beere}, \citenamefont {Jones},
  \citenamefont {Ritchie}, \citenamefont {Kelly},\ and\ \citenamefont
  {Smith}}]{AlTaie:2013ei}%
  \BibitemOpen
  \bibfield  {author} {\bibinfo {author} {\bibfnamefont {H.}~\bibnamefont
  {Al-Taie}}, \bibinfo {author} {\bibfnamefont {L.~W.}\ \bibnamefont {Smith}},
  \bibinfo {author} {\bibfnamefont {B.}~\bibnamefont {Xu}}, \bibinfo {author}
  {\bibfnamefont {P.}~\bibnamefont {See}}, \bibinfo {author} {\bibfnamefont
  {J.~P.}\ \bibnamefont {Griffiths}}, \bibinfo {author} {\bibfnamefont {H.~E.}\
  \bibnamefont {Beere}}, \bibinfo {author} {\bibfnamefont {G.~A.~C.}\
  \bibnamefont {Jones}}, \bibinfo {author} {\bibfnamefont {D.~A.}\ \bibnamefont
  {Ritchie}}, \bibinfo {author} {\bibfnamefont {M.~J.}\ \bibnamefont {Kelly}},
  \ and\ \bibinfo {author} {\bibfnamefont {C.~G.}\ \bibnamefont {Smith}},\
  }\href@noop {} {\bibfield  {journal} {\bibinfo  {journal} {Applied Physics
  Letters}\ }\textbf {\bibinfo {volume} {102}},\ \bibinfo {pages} {243102}
  (\bibinfo {year} {2013})}\BibitemShut {NoStop}%
\bibitem [{\citenamefont {Mourik}\ \emph {et~al.}(2012)\citenamefont {Mourik},
  \citenamefont {Zuo}, \citenamefont {Frolov}, \citenamefont {Plissard},
  \citenamefont {Bakkers},\ and\ \citenamefont {Kouwenhoven}}]{Mourik:2012je}%
  \BibitemOpen
  \bibfield  {author} {\bibinfo {author} {\bibfnamefont {V.}~\bibnamefont
  {Mourik}}, \bibinfo {author} {\bibfnamefont {K.}~\bibnamefont {Zuo}},
  \bibinfo {author} {\bibfnamefont {S.~M.}\ \bibnamefont {Frolov}}, \bibinfo
  {author} {\bibfnamefont {S.~R.}\ \bibnamefont {Plissard}}, \bibinfo {author}
  {\bibfnamefont {E.~P. A.~M.}\ \bibnamefont {Bakkers}}, \ and\ \bibinfo
  {author} {\bibfnamefont {L.~P.}\ \bibnamefont {Kouwenhoven}},\ }\href@noop {}
  {\bibfield  {journal} {\bibinfo  {journal} {Science}\ }\textbf {\bibinfo
  {volume} {336}},\ \bibinfo {pages} {1003} (\bibinfo {year}
  {2012})}\BibitemShut {NoStop}%
\bibitem [{\citenamefont {Das}\ \emph {et~al.}(2012)\citenamefont {Das},
  \citenamefont {Ronen}, \citenamefont {Most}, \citenamefont {Oreg},
  \citenamefont {Heiblum},\ and\ \citenamefont {Shtrikman}}]{Das:2012hi}%
  \BibitemOpen
  \bibfield  {author} {\bibinfo {author} {\bibfnamefont {A.}~\bibnamefont
  {Das}}, \bibinfo {author} {\bibfnamefont {Y.}~\bibnamefont {Ronen}}, \bibinfo
  {author} {\bibfnamefont {Y.}~\bibnamefont {Most}}, \bibinfo {author}
  {\bibfnamefont {Y.}~\bibnamefont {Oreg}}, \bibinfo {author} {\bibfnamefont
  {M.}~\bibnamefont {Heiblum}}, \ and\ \bibinfo {author} {\bibfnamefont
  {H.}~\bibnamefont {Shtrikman}},\ }\href@noop {} {\bibfield  {journal}
  {\bibinfo  {journal} {Nature Physics}\ }\textbf {\bibinfo {volume} {8}},\
  \bibinfo {pages} {887} (\bibinfo {year} {2012})}\BibitemShut {NoStop}%
\bibitem [{\citenamefont {Hassler}\ \emph {et~al.}(2011)\citenamefont
  {Hassler}, \citenamefont {Akhmerov},\ and\ \citenamefont
  {Beenakker}}]{Hassler:2011gj}%
  \BibitemOpen
  \bibfield  {author} {\bibinfo {author} {\bibfnamefont {F.}~\bibnamefont
  {Hassler}}, \bibinfo {author} {\bibfnamefont {A.~R.}\ \bibnamefont
  {Akhmerov}}, \ and\ \bibinfo {author} {\bibfnamefont {C.~W.~J.}\ \bibnamefont
  {Beenakker}},\ }\href@noop {} {\bibfield  {journal} {\bibinfo  {journal} {New
  Journal of Physics}\ }\textbf {\bibinfo {volume} {13}},\ \bibinfo {pages}
  {095004} (\bibinfo {year} {2011})}\BibitemShut {NoStop}%
\bibitem [{\citenamefont {Ginossar}\ and\ \citenamefont
  {Grosfeld}(2014)}]{Ginossar:1jd}%
  \BibitemOpen
  \bibfield  {author} {\bibinfo {author} {\bibfnamefont {E.}~\bibnamefont
  {Ginossar}}\ and\ \bibinfo {author} {\bibfnamefont {E.}~\bibnamefont
  {Grosfeld}},\ }\href@noop {} {\bibfield  {journal} {\bibinfo  {journal}
  {Nature Communications}\ }\textbf {\bibinfo {volume} {5}},\ \bibinfo {pages}
  {4772} (\bibinfo {year} {2014})}\BibitemShut {NoStop}%
\end{thebibliography}

%

\end{document}